\documentclass[osajnl,twocolumn,showpacs,superscriptaddress,10pt]{revtex4-1} 
\usepackage{amsmath,amssymb,graphicx}
\begin{document}

\title{Ultrafast laser inscription of mid-IR directional couplers for stellar interferometry}

\author{Alexander Arriola} \email{a.arriola@hw.ac.uk}
\author{Sebabrata Mukherjee}
\author{Debaditya Choudhury}
\affiliation{Institute of Photonics and Quantum Sciences (IPaQS), School of Engineering and Physical Sciences, Heriot-Watt University, Edinburgh, EH14 4AS, Scotland}

\author{Lucas Labadie}
\affiliation{I. Physikalisches Institut, Universit\"{a}t zu K\"{o}ln, Z\"{u}lpicher Str. 77, 50937 K\"{o}ln, Germany}

\author{Robert R. Thomson$^1$}

\begin{abstract} We report the ultrafast laser fabrication and mid-IR  characterization (3.39 $\mu$m) of four-port evanescent field directional couplers. The couplers were fabricated in a commercial gallium lanthanum sulphide glass substrate using sub-picosecond laser pulses of 1030 nm light. Straight waveguides inscribed using optimal fabrication parameters were found to exhibit propagation losses of $\sim$~0.8 dB$\cdot$cm$^{-1}$.  A series of couplers were inscribed with different interaction lengths, and we demonstrate power splitting ratios of between 8$\%$ and 99$\%$ for mid-IR light with a wavelength of 3.39 $\mu$m. These results clearly demonstrate that ultrafast laser inscription can be used to fabricate high quality evanescent field couplers for future applications in astronomical interferometry.
\end{abstract}

\maketitle 

Astronomical optical instruments that utilize photonic concepts \cite{Bland-Hawthorn2009}, such as integrated optical waveguides \cite{Davis1996, Ho2006, Labadie2011, Thomson2007, Jovanovic2012a, DAmico2014}, Bragg-gratings \cite{Marshall2006, Spaleniak2014} and laser frequency combs \cite{Jones2000, Cundiff2003}, have the potential to revolutionize astronomy in a plethora of different areas. One such area is stellar interferometry which, by combining the light gathered by multiple widely spaced telescopes, enables imaging of celestial objects with extremely high spatial resolution (sub milli-arcsecond \cite{Kloppenborg2010, Renard2010, Millour2011}), well beyond that realistically achievable using a single telescope.

To image an object using a telescope interferometer array is a highly non-trivial task. The collected light must first be transported to a central location using vacuum delay lines. A high-speed fringe tracker is also necessary to correct for path length changes due to atmospheric conditions. Once at the central location, the transported light must be coherently combined using a beam-combining instrument \cite{Monnier2003} which analyses the mutual coherence between the light from each pair of telescopes in the array. In 1996, \textit{Kern et al.} \cite{Kern1996} proposed that integrated optic (IO) waveguide technologies could provide an inherently stable and scalable platform for beam combination. Since then, IO beam combiners \cite{LeBouquin2011, Charles2012, Jovanovic2012a, Minardi2012, Arriola2013}  have been utilized in a variety of leading interferometric beam combination instruments, such as the PIONIER beam combination instrument for the VLTI \cite{LeBouquin2011}, which operates in the astronomical H- and K- bands, between 1.5 and 1.8 $\mu$m and 2.0 and 2.4 $\mu$m respectively.

One of the primary science drivers for the future development of stellar interferometry is exoplanetary science, since it should enable high resolution direct imaging of circumstellar disks and earth-like exoplanets. These cool objects predominantly emit blackbody radiation in the mid-infrared (mid-IR)  \cite{Labadie2009}, and for this reason the mid-IR wavelength band (i.e.  2.0 -12.0 $\mu$m) is a spectral region of particular importance to modern astronomy and stellar interferometry in particular. There is, therefore, a need to develop IO beam combination technologies for these longer mid-IR wavelengths. 

To date, prototype mid-IR IO beam combiners have been demonstrated using a variety of platforms, including Ti-indiffusion in LiNbO$_3$ \cite{Hsiao2009, Heidmann2011} and laser written waveguides on As$_2$Se$_3$ thin-films \cite{Labadie2011}. Both of these technologies are planar, however, which makes low cross-talk waveguide cross-overs extremely difficult. Furthermore, both of these technologies also result in optical waveguides with asymmetric refractive index profiles, a potential problem for high efficiency input coupling of light from an optical fibre.

With the drawbacks mentioned above in mind, we recently \cite{Rodenas2012} used ultrafast laser inscription (ULI) \cite{Davis1996} to fabricate a prototype 3-telescope mid-IR IO beam combination component. In contrast to other laser writing technologies, ULI relies on the non-linear absorption of focused sub-bandgap ultrashort pulses to locally modify the refractive index of the substrate material. By translating the material in three-dimensions through the laser focus, ULI enables the direct inscription of three-dimensional optical waveguide circuits with zero cross-talk cross-overs and highly symmetric refractive index profiles. In our earlier proof-of-concept work, however, the beam combination was performed using Y-junctions. These junctions are inherently lossy, and a useful IO beam-combiner must utilize evanescent field couplers rather than crude Y-junctions. In this work, we report the ULI fabrication and characterization of mid-IR evanescent field couplers for operation in the astronomical L-band (3.39 $\mu$m).

All structures reported in this letter were fabricated using a Menlo BlueCut fibre laser, which supplied $\sim$400 fs pulses of 1030 nm light at a pulse repetition frequency of 500 kHz. For our purposes, the polarization of the laser was adjusted to circular and focused approximately 200 $\mu$m below the surface of the substrate using a 0.4 NA lens. To translate the substrate (Gallium Lanthanum Sulphide (GLS) from ChG-Southampton) through the laser focus we used a set of computer controlled Aerotech air-bearing XYZ translation stages (ABL series). For the remainder of this paper, we will refer to the laser propagation axis inside the sample as the z-axis, the primary waveguide axis as the x-axis and the y-axis as the orthogonal axis to both the z- and x-axis. 

A ULI parameter investigation was performed, during which single straight waveguides were inscribed using pulse energies ranging from 80 to 15 nJ (in steps of 5$\%$ variation from the previous pulse energy), and translation velocities of 2, 4, 8 and 16 mm/s. The cross-section of the inscribed waveguides was controlled using the well-known multiscan waveguide shaping technique \cite{Nasu2005, Said2004, Thomson2011}. In this case, 21 scans with an inter-scan separation of 0.3 $\mu$m along the y-axis were used. After inscription, the end facets of the structures were ground and polished to an optical quality finish. The guiding properties of the inscribed structures were then investigated by coupling vertically polarized 3.39 $\mu$m light from a HeNe laser into the structures using free space coupling, while imaging the output of the structure onto a WinCam3D microbolometer beam profiler. This camera is sensitive in the wavelength range 2-16~$\mu$m. The optimal waveguides within the parameter range were found to be those fabricated using 70 nJ pulses (incident on the sample), a translation velocity of 4 mm/s.

\begin{figure}[htbp]
\centerline{\includegraphics[width=1\columnwidth]{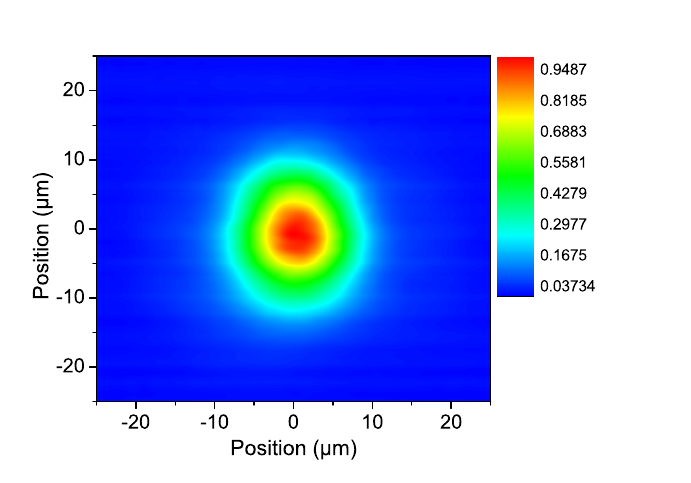}}
\caption{Mode field profile at 3.39 $\mu$m of the single-mode waveguide.}
\label{SM_MFD}
\end{figure}

As shown in Fig.~\ref{SM_MFD}, these waveguides were observed to support a single-mode at 3.39 $\mu$m, with a mode field diameter $\sim$22.43 $\pm$ 0.5 $\mu$m and $\sim$26.56  $\pm$ 0.5 $\mu$m in the y- and z-axis respectively (Fig.~\ref{SM_MFD}). The propagation loss of these optimum waveguides was measured using the cut-back technique and found to be $\sim$~0.8 $\pm$ 0.05 dB$\cdot$cm$^{-1}$. 

\begin{figure}[htbp]
\centerline{\includegraphics[width=1.05\columnwidth]{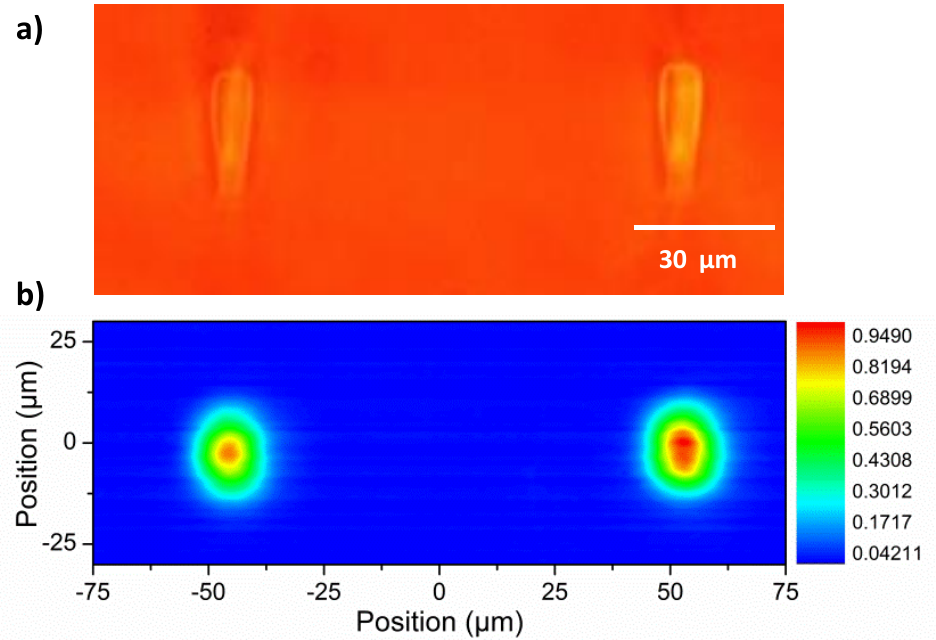}}
\caption{End-on bright-field microscope image of the two waveguides of a coupler inscribed in GLS glass using the multiscan technique(a), together with the two modes of one of the couplers (b).}
\label{Coupler_image}
\end{figure}

Using the optimal ULI parameters identified above, a series of evanescent field couplers were inscribed. Fig.~\ref{Coupler_image} shows a bright-field microscope image of the input facet of the two waveguides of the coupler fabricated in GLS glass (a), together with the two output modes of one of the couplers (b).

\begin{figure}[htbp]
\centerline{\includegraphics[width=0.9\columnwidth]{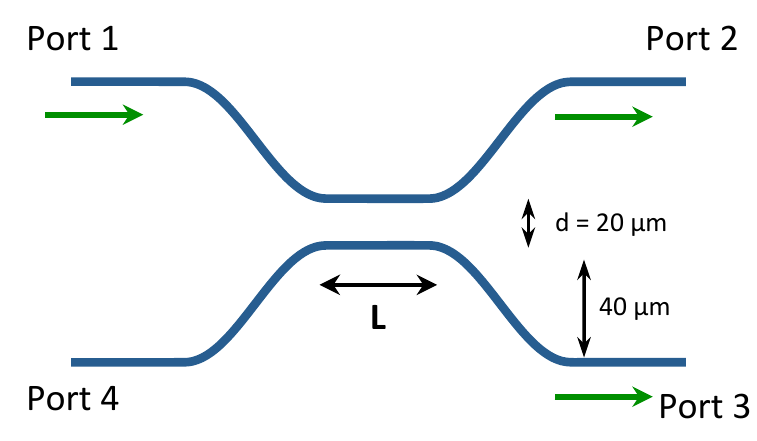}}
\caption{Schematic of the interference coupler structure, where d is the distance between the two waveguides along the interaction length (L).}
\label{Coupler_description}
\end{figure}

Each of the couplers were formed from two single-mode waveguides, and each of these waveguides consisted of straight lead-in and lead-out sections, bends that were inscribed had a sine-squared form ($y(x)~=~A~\cdot~\sin^2(\frac{2\pi x}{4L_c})$, with $A$ = 40 $\mu$m and  $L_c~=~$3 mm), and a straight interaction region of length ÒLÓ (Fig.~\ref{Coupler_description}). $A$ represents the change in the y-axis for the curve, while $L_c$ is the distance in the x-axis for this lateral displacement to occur. The lead-in and lead-out sections of the two single-mode waveguides for each coupler were separated by 100 $\mu$m to avoid evanescent coupling in these sections. For all couplers, the core-to-core separation in the interaction region was set to 20 $\mu$m. While keeping all other parameters constant, 32 couplers were inscribed, using interaction lengths between 2 and 10 mm in steps of 250 $\mu$m.

\begin{figure}[htbp]
\centerline{\includegraphics[width=1.1\columnwidth]{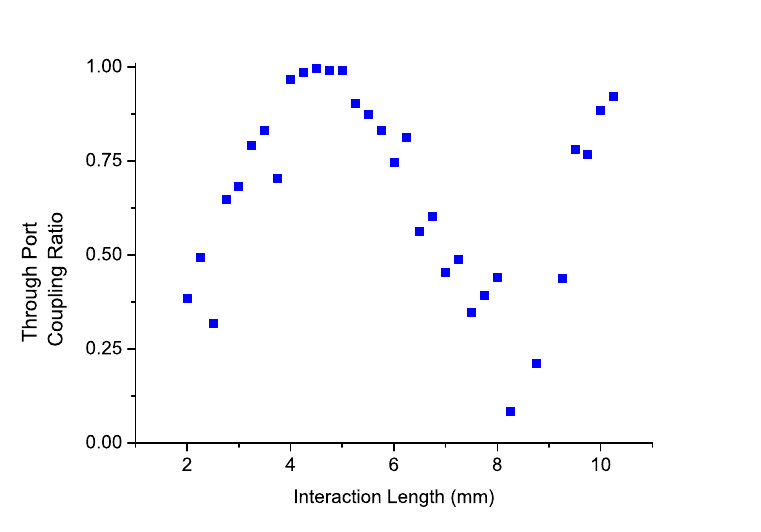}}
\caption{Through port coupling ratio dependence on the interaction length for a 20~$\mu$m separation.}
\label{Coupler_Graph}
\end{figure}

The power coupling ratios of each coupler were evaluated by measuring the Port 2/(Port 2+Port3) output power ratio when coupling vertically polarized 3.39 $\mu$m light into port 1. These measurements were made using a PbSe fixed-gain detector (1.5-4.8 $\mu$m) from Thorlabs. Figure~\ref{Coupler_Graph} shows the through port coupling ratio as a function of interaction length, where it can be seen that the measured coupling ratio varies between 8$\%$ and 99$\%$, with a beat length of $\sim$5.75 mm. It can also be seen from Fig.~\ref{Coupler_Graph} that the coupling ratio would be unlikely to reach 100$\%$, regardless of the sampling of the interaction length. This behaviour is characteristic of differences in the propagation constants of the two single mode waveguides that form the coupler \cite{Eaton2006}, a possibility that could be due to a variety of reasons. For example, it is possible that the optical properties (e.g. mode index) of the first waveguide for each coupler are directly affected but the inscription of the second waveguide - or indeed vice versa, or it is also plausible that temperature variations in the lab ($\sim\pm$ $1\,^{\circ}{\rm C}$) during the inscription process are sufficient to change the ULI parameters over the course of the experiment. 

We observed that the couplers exhibited the same total throughput as a straight single mode waveguide of the same length indicating that bend losses and radiation losses in the couplers were not signicant. Based on a $\sim$ 0.8 dB/cm $\pm$ 0.05 dB/cm propagation loss, as measured for the straight single mode waveguides, we conclude that the couplers would exhibit a total maximum throughput of loss of $\sim$ 1.6 dB/cm $\pm$ 0.1 dB/cm. This of course assumes zero Fresnel reections and input and output coupling losses.

In conclusion, we have demonstrated the ULI fabrication of prototype four-port evanescent field  couplers for the mid-IR region. The devices were tested using monochromatic 3.39 $\mu$m light, and were found to exhibit a clear power coupling dependency on interaction length, with a beat length of 5.75 mm. The weak dependency of the power-coupling ratio on interaction length will, in the future, enable precise fine tuning of the coupler characteristics for optimal operation in beam combination instruments. We believe that this demonstration, when combined with the 3D fabrication capability of ULI, will open the way towards highly stable and scalable beam combination instruments for mid-IR stellar interferometry and nulling interferometry.

We gratefully acknowledge funding from the UK Science and Technologies Facilities Council (STFC), in the form of a RRT«s STFC Advanced Fellowship (ST/H005595/1) and through the STFC Project Research and Development (STFC-PRD) scheme (ST/K00235X/1), and also from the Royal Society (RG110551). We also sthank the European Union for funding via the OPTICON Research Infrastructure for Optical/IR astronomy (EU-FP7 226604).


\begin{thebibliography}{0}%
\makeatletter
\providecommand \@ifxundefined [1]{%
 \@ifx{#1\undefined}
}%
\providecommand \@ifnum [1]{%
 \ifnum #1\expandafter \@firstoftwo
 \else \expandafter \@secondoftwo
 \fi
}%
\providecommand \@ifx [1]{%
 \ifx #1\expandafter \@firstoftwo
 \else \expandafter \@secondoftwo
 \fi
}%
\providecommand \natexlab [1]{#1}%
\providecommand \enquote  [1]{``#1''}%
\providecommand \bibnamefont  [1]{#1}%
\providecommand \bibfnamefont [1]{#1}%
\providecommand \citenamefont [1]{#1}%
\providecommand \href@noop [0]{\@secondoftwo}%
\providecommand \href [0]{\begingroup \@sanitize@url \@href}%
\providecommand \@href[1]{\@@startlink{#1}\@@href}%
\providecommand \@@href[1]{\endgroup#1\@@endlink}%
\providecommand \@sanitize@url [0]{\catcode `\\12\catcode `\$12\catcode
  `\&12\catcode `\#12\catcode `\^12\catcode `\_12\catcode `\%12\relax}%
\providecommand \@@startlink[1]{}%
\providecommand \@@endlink[0]{}%
\providecommand \url  [0]{\begingroup\@sanitize@url \@url }%
\providecommand \@url [1]{\endgroup\@href {#1}{\urlprefix }}%
\providecommand \urlprefix  [0]{URL }%
\providecommand \Eprint [0]{\href }%
\providecommand \doibase [0]{http://dx.doi.org/}%
\providecommand \selectlanguage [0]{\@gobble}%
\providecommand \bibinfo  [0]{\@secondoftwo}%
\providecommand \bibfield  [0]{\@secondoftwo}%
\providecommand \translation [1]{[#1]}%
\providecommand \BibitemOpen [0]{}%
\providecommand \bibitemStop [0]{}%
\providecommand \bibitemNoStop [0]{.\EOS\space}%
\providecommand \EOS [0]{\spacefactor3000\relax}%
\providecommand \BibitemShut  [1]{\csname bibitem#1\endcsname}%
\let\auto@bib@innerbib\@empty
\end{thebibliography}%


\begin{thebibliography}{28}%
\makeatletter
\providecommand \@ifxundefined [1]{%
 \@ifx{#1\undefined}
}%
\providecommand \@ifnum [1]{%
 \ifnum #1\expandafter \@firstoftwo
 \else \expandafter \@secondoftwo
 \fi
}%
\providecommand \@ifx [1]{%
 \ifx #1\expandafter \@firstoftwo
 \else \expandafter \@secondoftwo
 \fi
}%
\providecommand \natexlab [1]{#1}%
\providecommand \enquote  [1]{``#1''}%
\providecommand \bibnamefont  [1]{#1}%
\providecommand \bibfnamefont [1]{#1}%
\providecommand \citenamefont [1]{#1}%
\providecommand \href@noop [0]{\@secondoftwo}%
\providecommand \href [0]{\begingroup \@sanitize@url \@href}%
\providecommand \@href[1]{\@@startlink{#1}\@@href}%
\providecommand \@@href[1]{\endgroup#1\@@endlink}%
\providecommand \@sanitize@url [0]{\catcode `\\12\catcode `\$12\catcode
  `\&12\catcode `\#12\catcode `\^12\catcode `\_12\catcode `\%12\relax}%
\providecommand \@@startlink[1]{}%
\providecommand \@@endlink[0]{}%
\providecommand \url  [0]{\begingroup\@sanitize@url \@url }%
\providecommand \@url [1]{\endgroup\@href {#1}{\urlprefix }}%
\providecommand \urlprefix  [0]{URL }%
\providecommand \Eprint [0]{\href }%
\providecommand \doibase [0]{http://dx.doi.org/}%
\providecommand \selectlanguage [0]{\@gobble}%
\providecommand \bibinfo  [0]{\@secondoftwo}%
\providecommand \bibfield  [0]{\@secondoftwo}%
\providecommand \translation [1]{[#1]}%
\providecommand \BibitemOpen [0]{}%
\providecommand \bibitemStop [0]{}%
\providecommand \bibitemNoStop [0]{.\EOS\space}%
\providecommand \EOS [0]{\spacefactor3000\relax}%
\providecommand \BibitemShut  [1]{\csname bibitem#1\endcsname}%
\let\auto@bib@innerbib\@empty
\bibitem [{\citenamefont {Bland-Hawthorn}\ and\ \citenamefont
  {Kern}(2009)}]{Bland-Hawthorn2009}%
  \BibitemOpen
  \bibfield  {author} {\bibinfo {author} {\bibfnamefont {J.}~\bibnamefont
  {Bland-Hawthorn}}\ and\ \bibinfo {author} {\bibfnamefont {P.}~\bibnamefont
  {Kern}},\ }\href {http://www.ncbi.nlm.nih.gov/pubmed/19189019} {\bibfield
  {journal} {\bibinfo  {journal} {Optics Express}\ }\textbf {\bibinfo {volume}
  {17}},\ \bibinfo {pages} {1880} (\bibinfo {year} {2009})}\BibitemShut
  {NoStop}%
\bibitem [{\citenamefont {Davis}\ \emph {et~al.}(1996)\citenamefont {Davis},
  \citenamefont {Miura}, \citenamefont {Sugimoto},\ and\ \citenamefont
  {Hirao}}]{Davis1996}%
  \BibitemOpen
  \bibfield  {author} {\bibinfo {author} {\bibfnamefont {K.~M.}\ \bibnamefont
  {Davis}}, \bibinfo {author} {\bibfnamefont {K.}~\bibnamefont {Miura}},
  \bibinfo {author} {\bibfnamefont {N.}~\bibnamefont {Sugimoto}}, \ and\
  \bibinfo {author} {\bibfnamefont {K.}~\bibnamefont {Hirao}},\ }\href {<Go to
  ISI>://WOS:A1996VP37200008} {\bibfield  {journal} {\bibinfo  {journal}
  {Optics Letters}\ }\textbf {\bibinfo {volume} {21}},\ \bibinfo {pages} {1729}
  (\bibinfo {year} {1996})}\BibitemShut {NoStop}%
\bibitem [{\citenamefont {H\^{o}}\ \emph {et~al.}(2006)\citenamefont {H\^{o}},
  \citenamefont {Phillips}, \citenamefont {Qiao}, \citenamefont {Allen},
  \citenamefont {Krishnaswami}, \citenamefont {Riley}, \citenamefont {Myers},\
  and\ \citenamefont {Anheier}}]{Ho2006}%
  \BibitemOpen
  \bibfield  {author} {\bibinfo {author} {\bibfnamefont {N.}~\bibnamefont
  {H\^{o}}}, \bibinfo {author} {\bibfnamefont {M.~C.}\ \bibnamefont
  {Phillips}}, \bibinfo {author} {\bibfnamefont {H.}~\bibnamefont {Qiao}},
  \bibinfo {author} {\bibfnamefont {P.~J.}\ \bibnamefont {Allen}}, \bibinfo
  {author} {\bibfnamefont {K.}~\bibnamefont {Krishnaswami}}, \bibinfo {author}
  {\bibfnamefont {B.~J.}\ \bibnamefont {Riley}}, \bibinfo {author}
  {\bibfnamefont {T.~L.}\ \bibnamefont {Myers}}, \ and\ \bibinfo {author}
  {\bibfnamefont {N.~C.}\ \bibnamefont {Anheier}},\ }\href
  {http://www.ncbi.nlm.nih.gov/pubmed/16729095} {\bibfield  {journal} {\bibinfo
   {journal} {Optics Letters}\ }\textbf {\bibinfo {volume} {31}},\ \bibinfo
  {pages} {1860} (\bibinfo {year} {2006})}\BibitemShut {NoStop}%
\bibitem [{\citenamefont {Labadie}\ \emph {et~al.}(2011)\citenamefont
  {Labadie}, \citenamefont {Mart\'{\i}n}, \citenamefont {Anheier},
  \citenamefont {Arezki}, \citenamefont {Qiao}, \citenamefont {Bernacki},\ and\
  \citenamefont {Kern}}]{Labadie2011}%
  \BibitemOpen
  \bibfield  {author} {\bibinfo {author} {\bibfnamefont {L.}~\bibnamefont
  {Labadie}}, \bibinfo {author} {\bibfnamefont {G.}~\bibnamefont
  {Mart\'{\i}n}}, \bibinfo {author} {\bibfnamefont {N.~C.}\ \bibnamefont
  {Anheier}}, \bibinfo {author} {\bibfnamefont {B.}~\bibnamefont {Arezki}},
  \bibinfo {author} {\bibfnamefont {H.~A.}\ \bibnamefont {Qiao}}, \bibinfo
  {author} {\bibfnamefont {B.}~\bibnamefont {Bernacki}}, \ and\ \bibinfo
  {author} {\bibfnamefont {P.}~\bibnamefont {Kern}},\ }\href@noop {} {\bibfield
   {journal} {\bibinfo  {journal} {Astronomy and Astrophysics}\ }\textbf
  {\bibinfo {volume} {48}},\ \bibinfo {pages} {1} (\bibinfo {year}
  {2011})}\BibitemShut {NoStop}%
\bibitem [{\citenamefont {Thomson}\ \emph {et~al.}(2007)\citenamefont
  {Thomson}, \citenamefont {Bookey}, \citenamefont {Psaila}, \citenamefont
  {Fender}, \citenamefont {Campbell}, \citenamefont {Macpherson}, \citenamefont
  {Barton}, \citenamefont {Reid},\ and\ \citenamefont {Kar}}]{Thomson2007}%
  \BibitemOpen
  \bibfield  {author} {\bibinfo {author} {\bibfnamefont {R.~R.}\ \bibnamefont
  {Thomson}}, \bibinfo {author} {\bibfnamefont {H.~T.}\ \bibnamefont {Bookey}},
  \bibinfo {author} {\bibfnamefont {N.~D.}\ \bibnamefont {Psaila}}, \bibinfo
  {author} {\bibfnamefont {A.}~\bibnamefont {Fender}}, \bibinfo {author}
  {\bibfnamefont {S.}~\bibnamefont {Campbell}}, \bibinfo {author}
  {\bibfnamefont {W.~N.}\ \bibnamefont {Macpherson}}, \bibinfo {author}
  {\bibfnamefont {J.~S.}\ \bibnamefont {Barton}}, \bibinfo {author}
  {\bibfnamefont {D.~T.}\ \bibnamefont {Reid}}, \ and\ \bibinfo {author}
  {\bibfnamefont {A.~K.}\ \bibnamefont {Kar}},\ }\href
  {http://www.ncbi.nlm.nih.gov/pubmed/19547529} {\bibfield  {journal} {\bibinfo
   {journal} {Optics Express}\ }\textbf {\bibinfo {volume} {15}},\ \bibinfo
  {pages} {11691} (\bibinfo {year} {2007})}\BibitemShut {NoStop}%
\bibitem [{\citenamefont {Jovanovic}\ \emph {et~al.}(2012)\citenamefont
  {Jovanovic}, \citenamefont {Tuthill}, \citenamefont {Norris}, \citenamefont
  {Gross}, \citenamefont {Stewart}, \citenamefont {Charles}, \citenamefont
  {Lacour}, \citenamefont {Ams}, \citenamefont {Lawrence}, \citenamefont
  {Lehmann}, \citenamefont {Niel}, \citenamefont {Robertson}, \citenamefont
  {Marshall}, \citenamefont {Ireland}, \citenamefont {Fuerbach},\ and\
  \citenamefont {Withford}}]{Jovanovic2012a}%
  \BibitemOpen
  \bibfield  {author} {\bibinfo {author} {\bibfnamefont {N.}~\bibnamefont
  {Jovanovic}}, \bibinfo {author} {\bibfnamefont {P.~G.}\ \bibnamefont
  {Tuthill}}, \bibinfo {author} {\bibfnamefont {B.}~\bibnamefont {Norris}},
  \bibinfo {author} {\bibfnamefont {S.}~\bibnamefont {Gross}}, \bibinfo
  {author} {\bibfnamefont {P.}~\bibnamefont {Stewart}}, \bibinfo {author}
  {\bibfnamefont {N.}~\bibnamefont {Charles}}, \bibinfo {author} {\bibfnamefont
  {S.}~\bibnamefont {Lacour}}, \bibinfo {author} {\bibfnamefont
  {M.}~\bibnamefont {Ams}}, \bibinfo {author} {\bibfnamefont {J.~S.}\
  \bibnamefont {Lawrence}}, \bibinfo {author} {\bibfnamefont {A.}~\bibnamefont
  {Lehmann}}, \bibinfo {author} {\bibfnamefont {C.}~\bibnamefont {Niel}},
  \bibinfo {author} {\bibfnamefont {J.~G.}\ \bibnamefont {Robertson}}, \bibinfo
  {author} {\bibfnamefont {G.~D.}\ \bibnamefont {Marshall}}, \bibinfo {author}
  {\bibfnamefont {M.}~\bibnamefont {Ireland}}, \bibinfo {author} {\bibfnamefont
  {A.}~\bibnamefont {Fuerbach}}, \ and\ \bibinfo {author} {\bibfnamefont
  {M.~J.}\ \bibnamefont {Withford}},\ }\href@noop {} {\bibfield  {journal}
  {\bibinfo  {journal} {Monthly Notices of the Royal Astronomical Society}\
  }\textbf {\bibinfo {volume} {427}},\ \bibinfo {pages} {10} (\bibinfo {year}
  {2012})}\BibitemShut {NoStop}%
\bibitem [{\citenamefont {DÕAmico}\ \emph {et~al.}(2014)\citenamefont
  {DÕAmico}, \citenamefont {Cheng}, \citenamefont {Mauclair}, \citenamefont
  {Troles}, \citenamefont {Calvez}, \citenamefont {Nazabal}, \citenamefont
  {Caillaud}, \citenamefont {Martin}, \citenamefont {Arezki}, \citenamefont
  {LeCoarer}, \citenamefont {Kern},\ and\ \citenamefont {Stoian}}]{DAmico2014}%
  \BibitemOpen
  \bibfield  {author} {\bibinfo {author} {\bibfnamefont {C.}~\bibnamefont
  {DÕAmico}}, \bibinfo {author} {\bibfnamefont {G.}~\bibnamefont {Cheng}},
  \bibinfo {author} {\bibfnamefont {C.}~\bibnamefont {Mauclair}}, \bibinfo
  {author} {\bibfnamefont {J.}~\bibnamefont {Troles}}, \bibinfo {author}
  {\bibfnamefont {L.}~\bibnamefont {Calvez}}, \bibinfo {author} {\bibfnamefont
  {V.}~\bibnamefont {Nazabal}}, \bibinfo {author} {\bibfnamefont
  {C.}~\bibnamefont {Caillaud}}, \bibinfo {author} {\bibfnamefont
  {G.}~\bibnamefont {Martin}}, \bibinfo {author} {\bibfnamefont
  {B.}~\bibnamefont {Arezki}}, \bibinfo {author} {\bibfnamefont
  {E.}~\bibnamefont {LeCoarer}}, \bibinfo {author} {\bibfnamefont
  {P.}~\bibnamefont {Kern}}, \ and\ \bibinfo {author} {\bibfnamefont
  {R.}~\bibnamefont {Stoian}},\ }\href {\doibase 10.1364/OE.22.013091}
  {\bibfield  {journal} {\bibinfo  {journal} {Optics Express}\ }\textbf
  {\bibinfo {volume} {22}},\ \bibinfo {pages} {13091} (\bibinfo {year}
  {2014})}\BibitemShut {NoStop}%
\bibitem [{\citenamefont {Marshall}\ \emph {et~al.}(2006)\citenamefont
  {Marshall}, \citenamefont {Ams},\ and\ \citenamefont
  {Withford}}]{Marshall2006}%
  \BibitemOpen
  \bibfield  {author} {\bibinfo {author} {\bibfnamefont {G.~D.}\ \bibnamefont
  {Marshall}}, \bibinfo {author} {\bibfnamefont {M.}~\bibnamefont {Ams}}, \
  and\ \bibinfo {author} {\bibfnamefont {M.~J.}\ \bibnamefont {Withford}},\
  }\href {http://www.ncbi.nlm.nih.gov/pubmed/16936859} {\bibfield  {journal}
  {\bibinfo  {journal} {Optics Letters}\ }\textbf {\bibinfo {volume} {31}},\
  \bibinfo {pages} {2690} (\bibinfo {year} {2006})}\BibitemShut {NoStop}%
\bibitem [{\citenamefont {Spaleniak}\ \emph {et~al.}(2014)\citenamefont
  {Spaleniak}, \citenamefont {Gross}, \citenamefont {Jovanovic}, \citenamefont
  {Williams}, \citenamefont {Lawrence}, \citenamefont {Ireland},\ and\
  \citenamefont {Withford}}]{Spaleniak2014}%
  \BibitemOpen
  \bibfield  {author} {\bibinfo {author} {\bibfnamefont {I.}~\bibnamefont
  {Spaleniak}}, \bibinfo {author} {\bibfnamefont {S.}~\bibnamefont {Gross}},
  \bibinfo {author} {\bibfnamefont {N.}~\bibnamefont {Jovanovic}}, \bibinfo
  {author} {\bibfnamefont {R.~J.}\ \bibnamefont {Williams}}, \bibinfo {author}
  {\bibfnamefont {J.~S.}\ \bibnamefont {Lawrence}}, \bibinfo {author}
  {\bibfnamefont {M.~J.}\ \bibnamefont {Ireland}}, \ and\ \bibinfo {author}
  {\bibfnamefont {M.~J.}\ \bibnamefont {Withford}},\ }\href@noop {} {\bibfield
  {journal} {\bibinfo  {journal} {Laser \& Photonics Reviews}\ }\textbf
  {\bibinfo {volume} {8}},\ \bibinfo {pages} {L1} (\bibinfo {year}
  {2014})}\BibitemShut {NoStop}%
\bibitem [{\citenamefont {Jones}\ \emph {et~al.}(2000)\citenamefont {Jones},
  \citenamefont {Diddams}, \citenamefont {Ranka}, \citenamefont {Stentz},
  \citenamefont {Windeler}, \citenamefont {Hall},\ and\ \citenamefont
  {Cundiff}}]{Jones2000}%
  \BibitemOpen
  \bibfield  {author} {\bibinfo {author} {\bibfnamefont {D.~J.}\ \bibnamefont
  {Jones}}, \bibinfo {author} {\bibfnamefont {S.~A.}\ \bibnamefont {Diddams}},
  \bibinfo {author} {\bibfnamefont {J.~K.}\ \bibnamefont {Ranka}}, \bibinfo
  {author} {\bibfnamefont {A.}~\bibnamefont {Stentz}}, \bibinfo {author}
  {\bibfnamefont {R.~S.}\ \bibnamefont {Windeler}}, \bibinfo {author}
  {\bibfnamefont {J.~L.}\ \bibnamefont {Hall}}, \ and\ \bibinfo {author}
  {\bibfnamefont {S.~T.}\ \bibnamefont {Cundiff}},\ }\href {\doibase
  10.1126/science.288.5466.635} {\bibfield  {journal} {\bibinfo  {journal}
  {Science}\ }\textbf {\bibinfo {volume} {288}},\ \bibinfo {pages} {635}
  (\bibinfo {year} {2000})}\BibitemShut {NoStop}%
\bibitem [{\citenamefont {Cundiff}\ and\ \citenamefont
  {Ye}(2003)}]{Cundiff2003}%
  \BibitemOpen
  \bibfield  {author} {\bibinfo {author} {\bibfnamefont {S.~T.}\ \bibnamefont
  {Cundiff}}\ and\ \bibinfo {author} {\bibfnamefont {J.}~\bibnamefont {Ye}},\
  }\href@noop {} {\bibfield  {journal} {\bibinfo  {journal} {Reviews of Modern
  Physics}\ }\textbf {\bibinfo {volume} {75}},\ \bibinfo {pages} {325}
  (\bibinfo {year} {2003})}\BibitemShut {NoStop}%
\bibitem [{\citenamefont {Kloppenborg}\ \emph {et~al.}(2010)\citenamefont
  {Kloppenborg}, \citenamefont {Stencel}, \citenamefont {Monnier},
  \citenamefont {Schaefer}, \citenamefont {Zhao}, \citenamefont {Baron},
  \citenamefont {McAlister}, \citenamefont {{Ten Brummelaar}}, \citenamefont
  {Che}, \citenamefont {Farrington}, \citenamefont {Pedretti}, \citenamefont
  {Sallave-Goldfinger}, \citenamefont {Sturmann}, \citenamefont {Sturmann},
  \citenamefont {Thureau}, \citenamefont {Turner},\ and\ \citenamefont
  {Carroll}}]{Kloppenborg2010}%
  \BibitemOpen
  \bibfield  {author} {\bibinfo {author} {\bibfnamefont {B.}~\bibnamefont
  {Kloppenborg}}, \bibinfo {author} {\bibfnamefont {R.}~\bibnamefont
  {Stencel}}, \bibinfo {author} {\bibfnamefont {J.~D.}\ \bibnamefont
  {Monnier}}, \bibinfo {author} {\bibfnamefont {G.}~\bibnamefont {Schaefer}},
  \bibinfo {author} {\bibfnamefont {M.}~\bibnamefont {Zhao}}, \bibinfo {author}
  {\bibfnamefont {F.}~\bibnamefont {Baron}}, \bibinfo {author} {\bibfnamefont
  {H.}~\bibnamefont {McAlister}}, \bibinfo {author} {\bibfnamefont
  {T.}~\bibnamefont {{Ten Brummelaar}}}, \bibinfo {author} {\bibfnamefont
  {X.}~\bibnamefont {Che}}, \bibinfo {author} {\bibfnamefont {C.}~\bibnamefont
  {Farrington}}, \bibinfo {author} {\bibfnamefont {E.}~\bibnamefont
  {Pedretti}}, \bibinfo {author} {\bibfnamefont {P.~J.}\ \bibnamefont
  {Sallave-Goldfinger}}, \bibinfo {author} {\bibfnamefont {J.}~\bibnamefont
  {Sturmann}}, \bibinfo {author} {\bibfnamefont {L.}~\bibnamefont {Sturmann}},
  \bibinfo {author} {\bibfnamefont {N.}~\bibnamefont {Thureau}}, \bibinfo
  {author} {\bibfnamefont {N.}~\bibnamefont {Turner}}, \ and\ \bibinfo {author}
  {\bibfnamefont {S.~M.}\ \bibnamefont {Carroll}},\ }\href {\doibase
  10.1038/nature08968} {\bibfield  {journal} {\bibinfo  {journal} {Nature}\
  }\textbf {\bibinfo {volume} {464}},\ \bibinfo {pages} {870} (\bibinfo {year}
  {2010})}\BibitemShut {NoStop}%
\bibitem [{\citenamefont {Renard}\ \emph {et~al.}(2010)\citenamefont {Renard},
  \citenamefont {Malbet}, \citenamefont {Benisty}, \citenamefont
  {Thi\'{e}baut},\ and\ \citenamefont {Berger}}]{Renard2010}%
  \BibitemOpen
  \bibfield  {author} {\bibinfo {author} {\bibfnamefont {S.}~\bibnamefont
  {Renard}}, \bibinfo {author} {\bibfnamefont {F.}~\bibnamefont {Malbet}},
  \bibinfo {author} {\bibfnamefont {M.}~\bibnamefont {Benisty}}, \bibinfo
  {author} {\bibfnamefont {E.}~\bibnamefont {Thi\'{e}baut}}, \ and\ \bibinfo
  {author} {\bibfnamefont {J.}~\bibnamefont {Berger}},\ }\href@noop {}
  {\bibfield  {journal} {\bibinfo  {journal} {Astronomy and Astrophysics}\
  }\textbf {\bibinfo {volume} {519}},\ \bibinfo {pages} {1} (\bibinfo {year}
  {2010})}\BibitemShut {NoStop}%
\bibitem [{\citenamefont {Millour}\ \emph {et~al.}(2011)\citenamefont
  {Millour}, \citenamefont {Meilland}, \citenamefont {Chesneau}, \citenamefont
  {Stee}, \citenamefont {Kanaan}, \citenamefont {Petrov}, \citenamefont
  {Mourard},\ and\ \citenamefont {Kraus}}]{Millour2011}%
  \BibitemOpen
  \bibfield  {author} {\bibinfo {author} {\bibfnamefont {F.}~\bibnamefont
  {Millour}}, \bibinfo {author} {\bibfnamefont {A.}~\bibnamefont {Meilland}},
  \bibinfo {author} {\bibfnamefont {O.}~\bibnamefont {Chesneau}}, \bibinfo
  {author} {\bibfnamefont {P.}~\bibnamefont {Stee}}, \bibinfo {author}
  {\bibfnamefont {S.}~\bibnamefont {Kanaan}}, \bibinfo {author} {\bibfnamefont
  {R.}~\bibnamefont {Petrov}}, \bibinfo {author} {\bibfnamefont
  {D.}~\bibnamefont {Mourard}}, \ and\ \bibinfo {author} {\bibfnamefont
  {S.}~\bibnamefont {Kraus}},\ }\href@noop {} {\bibfield  {journal} {\bibinfo
  {journal} {Astronomy and Astrophysics}\ }\textbf {\bibinfo {volume} {526}},\
  \bibinfo {pages} {4} (\bibinfo {year} {2011})}\BibitemShut {NoStop}%
\bibitem [{\citenamefont {Monnier}(2003)}]{Monnier2003}%
  \BibitemOpen
  \bibfield  {author} {\bibinfo {author} {\bibfnamefont {J.~D.}\ \bibnamefont
  {Monnier}},\ }\href@noop {} {\bibfield  {journal} {\bibinfo  {journal}
  {Reports on Progress in Physics}\ }\textbf {\bibinfo {volume} {66}},\
  \bibinfo {pages} {789} (\bibinfo {year} {2003})}\BibitemShut {NoStop}%
\bibitem [{\citenamefont {Kern}\ and\ \citenamefont {Malbet}(1996)}]{Kern1996}%
  \BibitemOpen
  \bibfield  {author} {\bibinfo {author} {\bibfnamefont {P.}~\bibnamefont
  {Kern}}\ and\ \bibinfo {author} {\bibfnamefont {F.}~\bibnamefont {Malbet}},\
  }in\ \href@noop {} {\emph {\bibinfo {booktitle} {AstroFibÕ96}}}\ (\bibinfo
  {year} {1996})\BibitemShut {NoStop}%
\bibitem [{\citenamefont {{Le Bouquin}}\ \emph {et~al.}(2011)\citenamefont {{Le
  Bouquin}}, \citenamefont {Berger}, \citenamefont {Lazareff}, \citenamefont
  {Zins}, \citenamefont {Haguenauer}, \citenamefont {Jocou}, \citenamefont
  {P.Kern}, \citenamefont {Millan-Gabet}, \citenamefont {Traub}, \citenamefont
  {Absil}, \citenamefont {Augereau}, \citenamefont {Benisty}, \citenamefont
  {Blind}, \citenamefont {X.Bonfils}, \citenamefont {Bourget}, \citenamefont
  {Delboulbe}, \citenamefont {Feautrier}, \citenamefont {Germain},
  \citenamefont {Gitton}, \citenamefont {Gillier}, \citenamefont
  {M.Kiekebusch}, \citenamefont {J.Kluska}, \citenamefont {Knudstrup},
  \citenamefont {Labeye}, \citenamefont {Lizon}, \citenamefont {Monin},
  \citenamefont {Magnard}, \citenamefont {Malbet}, \citenamefont {Maurel},
  \citenamefont {M\'{e}nard}, \citenamefont {Micallef}, \citenamefont
  {Michaud}, \citenamefont {Montagnier}, \citenamefont {Morel}, \citenamefont
  {Moulin}, \citenamefont {Perraut}, \citenamefont {Popovic}, \citenamefont
  {Rabou}, \citenamefont {Rochat}, \citenamefont {C.Rojas}, \citenamefont
  {Roussel}, \citenamefont {Roux}, \citenamefont {Stadler}, \citenamefont
  {Stefl}, \citenamefont {Tatulli},\ and\ \citenamefont
  {Ventura}}]{LeBouquin2011}%
  \BibitemOpen
  \bibfield  {author} {\bibinfo {author} {\bibfnamefont {J.-B.}\ \bibnamefont
  {{Le Bouquin}}}, \bibinfo {author} {\bibfnamefont {J.-P.}\ \bibnamefont
  {Berger}}, \bibinfo {author} {\bibfnamefont {B.}~\bibnamefont {Lazareff}},
  \bibinfo {author} {\bibfnamefont {G.}~\bibnamefont {Zins}}, \bibinfo {author}
  {\bibfnamefont {P.}~\bibnamefont {Haguenauer}}, \bibinfo {author}
  {\bibfnamefont {L.}~\bibnamefont {Jocou}}, \bibinfo {author} {\bibnamefont
  {P.Kern}}, \bibinfo {author} {\bibfnamefont {R.}~\bibnamefont
  {Millan-Gabet}}, \bibinfo {author} {\bibfnamefont {W.}~\bibnamefont {Traub}},
  \bibinfo {author} {\bibfnamefont {O.}~\bibnamefont {Absil}}, \bibinfo
  {author} {\bibfnamefont {J.-C.}\ \bibnamefont {Augereau}}, \bibinfo {author}
  {\bibfnamefont {M.}~\bibnamefont {Benisty}}, \bibinfo {author} {\bibfnamefont
  {N.}~\bibnamefont {Blind}}, \bibinfo {author} {\bibnamefont {X.Bonfils}},
  \bibinfo {author} {\bibfnamefont {P.}~\bibnamefont {Bourget}}, \bibinfo
  {author} {\bibfnamefont {A.}~\bibnamefont {Delboulbe}}, \bibinfo {author}
  {\bibfnamefont {P.}~\bibnamefont {Feautrier}}, \bibinfo {author}
  {\bibfnamefont {M.}~\bibnamefont {Germain}}, \bibinfo {author} {\bibfnamefont
  {P.}~\bibnamefont {Gitton}}, \bibinfo {author} {\bibfnamefont
  {D.}~\bibnamefont {Gillier}}, \bibinfo {author} {\bibnamefont
  {M.Kiekebusch}}, \bibinfo {author} {\bibnamefont {J.Kluska}}, \bibinfo
  {author} {\bibfnamefont {J.}~\bibnamefont {Knudstrup}}, \bibinfo {author}
  {\bibfnamefont {P.}~\bibnamefont {Labeye}}, \bibinfo {author} {\bibfnamefont
  {J.-L.}\ \bibnamefont {Lizon}}, \bibinfo {author} {\bibfnamefont {J.-L.}\
  \bibnamefont {Monin}}, \bibinfo {author} {\bibfnamefont {Y.}~\bibnamefont
  {Magnard}}, \bibinfo {author} {\bibfnamefont {F.}~\bibnamefont {Malbet}},
  \bibinfo {author} {\bibfnamefont {D.}~\bibnamefont {Maurel}}, \bibinfo
  {author} {\bibfnamefont {F.}~\bibnamefont {M\'{e}nard}}, \bibinfo {author}
  {\bibfnamefont {M.}~\bibnamefont {Micallef}}, \bibinfo {author}
  {\bibfnamefont {L.}~\bibnamefont {Michaud}}, \bibinfo {author} {\bibfnamefont
  {G.}~\bibnamefont {Montagnier}}, \bibinfo {author} {\bibfnamefont
  {S.}~\bibnamefont {Morel}}, \bibinfo {author} {\bibfnamefont
  {T.}~\bibnamefont {Moulin}}, \bibinfo {author} {\bibfnamefont
  {K.}~\bibnamefont {Perraut}}, \bibinfo {author} {\bibfnamefont
  {D.}~\bibnamefont {Popovic}}, \bibinfo {author} {\bibfnamefont
  {P.}~\bibnamefont {Rabou}}, \bibinfo {author} {\bibfnamefont
  {S.}~\bibnamefont {Rochat}}, \bibinfo {author} {\bibnamefont {C.Rojas}},
  \bibinfo {author} {\bibfnamefont {F.}~\bibnamefont {Roussel}}, \bibinfo
  {author} {\bibfnamefont {A.}~\bibnamefont {Roux}}, \bibinfo {author}
  {\bibfnamefont {E.}~\bibnamefont {Stadler}}, \bibinfo {author} {\bibfnamefont
  {S.}~\bibnamefont {Stefl}}, \bibinfo {author} {\bibfnamefont
  {E.}~\bibnamefont {Tatulli}}, \ and\ \bibinfo {author} {\bibfnamefont
  {N.}~\bibnamefont {Ventura}},\ }\href@noop {} {\bibfield  {journal} {\bibinfo
   {journal} {Astronomy and Astrophysics}\ }\textbf {\bibinfo {volume} {535}},\
  \bibinfo {pages} {1} (\bibinfo {year} {2011})}\BibitemShut {NoStop}%
\bibitem [{\citenamefont {Charles}\ \emph {et~al.}(2012)\citenamefont
  {Charles}, \citenamefont {Jovanovic}, \citenamefont {Gross}, \citenamefont
  {Stewart}, \citenamefont {Norris}, \citenamefont {O'Byrne}, \citenamefont
  {Lawrence}, \citenamefont {Withford},\ and\ \citenamefont
  {Tuthill}}]{Charles2012}%
  \BibitemOpen
  \bibfield  {author} {\bibinfo {author} {\bibfnamefont {N.}~\bibnamefont
  {Charles}}, \bibinfo {author} {\bibfnamefont {N.}~\bibnamefont {Jovanovic}},
  \bibinfo {author} {\bibfnamefont {S.}~\bibnamefont {Gross}}, \bibinfo
  {author} {\bibfnamefont {P.}~\bibnamefont {Stewart}}, \bibinfo {author}
  {\bibfnamefont {B.}~\bibnamefont {Norris}}, \bibinfo {author} {\bibfnamefont
  {J.}~\bibnamefont {O'Byrne}}, \bibinfo {author} {\bibfnamefont {J.~S.}\
  \bibnamefont {Lawrence}}, \bibinfo {author} {\bibfnamefont {M.~J.}\
  \bibnamefont {Withford}}, \ and\ \bibinfo {author} {\bibfnamefont {P.~G.}\
  \bibnamefont {Tuthill}},\ }\href {<Go to ISI>://CCC:000309168100008
  http://www.osa.org} {\bibfield  {journal} {\bibinfo  {journal} {Applied
  Optics}\ }\textbf {\bibinfo {volume} {51}},\ \bibinfo {pages} {6489}
  (\bibinfo {year} {2012})}\BibitemShut {NoStop}%
\bibitem [{\citenamefont {Minardi}\ \emph {et~al.}(2012)\citenamefont
  {Minardi}, \citenamefont {Dreisow}, \citenamefont {Gr\"{a}fe}, \citenamefont
  {Nolte},\ and\ \citenamefont {Pertsch}}]{Minardi2012}%
  \BibitemOpen
  \bibfield  {author} {\bibinfo {author} {\bibfnamefont {S.}~\bibnamefont
  {Minardi}}, \bibinfo {author} {\bibfnamefont {F.}~\bibnamefont {Dreisow}},
  \bibinfo {author} {\bibfnamefont {M.}~\bibnamefont {Gr\"{a}fe}}, \bibinfo
  {author} {\bibfnamefont {S.}~\bibnamefont {Nolte}}, \ and\ \bibinfo {author}
  {\bibfnamefont {T.}~\bibnamefont {Pertsch}},\ }\href
  {http://www.ncbi.nlm.nih.gov/pubmed/22859075} {\bibfield  {journal} {\bibinfo
   {journal} {Optics Letters}\ }\textbf {\bibinfo {volume} {37}},\ \bibinfo
  {pages} {3030} (\bibinfo {year} {2012})}\BibitemShut {NoStop}%
\bibitem [{\citenamefont {Arriola}\ \emph {et~al.}(2013)\citenamefont
  {Arriola}, \citenamefont {Gross}, \citenamefont {Jovanovic}, \citenamefont
  {Charles}, \citenamefont {Tuthill}, \citenamefont {Olaizola}, \citenamefont
  {Fuerbach},\ and\ \citenamefont {Withford}}]{Arriola2013}%
  \BibitemOpen
  \bibfield  {author} {\bibinfo {author} {\bibfnamefont {A.}~\bibnamefont
  {Arriola}}, \bibinfo {author} {\bibfnamefont {S.}~\bibnamefont {Gross}},
  \bibinfo {author} {\bibfnamefont {N.}~\bibnamefont {Jovanovic}}, \bibinfo
  {author} {\bibfnamefont {N.}~\bibnamefont {Charles}}, \bibinfo {author}
  {\bibfnamefont {P.~G.}\ \bibnamefont {Tuthill}}, \bibinfo {author}
  {\bibfnamefont {S.~M.}\ \bibnamefont {Olaizola}}, \bibinfo {author}
  {\bibfnamefont {A.}~\bibnamefont {Fuerbach}}, \ and\ \bibinfo {author}
  {\bibfnamefont {M.~J.}\ \bibnamefont {Withford}},\ }\href {\doibase
  10.1038/ncomms1570.} {\bibfield  {journal} {\bibinfo  {journal} {Optics
  Express}\ }\textbf {\bibinfo {volume} {21}},\ \bibinfo {pages} {2978}
  (\bibinfo {year} {2013})}\BibitemShut {NoStop}%
\bibitem [{\citenamefont {Labadie}\ and\ \citenamefont
  {Wallner}(2009)}]{Labadie2009}%
  \BibitemOpen
  \bibfield  {author} {\bibinfo {author} {\bibfnamefont {L.}~\bibnamefont
  {Labadie}}\ and\ \bibinfo {author} {\bibfnamefont {O.}~\bibnamefont
  {Wallner}},\ }\href {http://www.ncbi.nlm.nih.gov/pubmed/19189025} {\bibfield
  {journal} {\bibinfo  {journal} {Optics Express}\ }\textbf {\bibinfo {volume}
  {17}},\ \bibinfo {pages} {1947} (\bibinfo {year} {2009})}\BibitemShut
  {NoStop}%
\bibitem [{\citenamefont {Hsiao}\ \emph {et~al.}(2009)\citenamefont {Hsiao},
  \citenamefont {Winick}, \citenamefont {Monnier},\ and\ \citenamefont
  {Berger}}]{Hsiao2009}%
  \BibitemOpen
  \bibfield  {author} {\bibinfo {author} {\bibfnamefont {H.-K.}\ \bibnamefont
  {Hsiao}}, \bibinfo {author} {\bibfnamefont {K.~A.}\ \bibnamefont {Winick}},
  \bibinfo {author} {\bibfnamefont {J.~D.}\ \bibnamefont {Monnier}}, \ and\
  \bibinfo {author} {\bibfnamefont {J.-P.}\ \bibnamefont {Berger}},\
  }\href@noop {} {\bibfield  {journal} {\bibinfo  {journal} {Optics Express}\
  }\textbf {\bibinfo {volume} {17}},\ \bibinfo {pages} {18489} (\bibinfo {year}
  {2009})}\BibitemShut {NoStop}%
\bibitem [{\citenamefont {Heidmann}\ \emph {et~al.}(2011)\citenamefont
  {Heidmann}, \citenamefont {Courjal},\ and\ \citenamefont
  {Martin}}]{Heidmann2011}%
  \BibitemOpen
  \bibfield  {author} {\bibinfo {author} {\bibfnamefont {S.}~\bibnamefont
  {Heidmann}}, \bibinfo {author} {\bibfnamefont {N.}~\bibnamefont {Courjal}}, \
  and\ \bibinfo {author} {\bibfnamefont {G.}~\bibnamefont {Martin}},\ }in\
  \href@noop {} {\emph {\bibinfo {booktitle} {European Quantum Electronics
  Conference}}}\ (\bibinfo {year} {2011})\BibitemShut {NoStop}%
\bibitem [{\citenamefont {R\'{o}denas}\ \emph {et~al.}(2012)\citenamefont
  {R\'{o}denas}, \citenamefont {Martin}, \citenamefont {Arezki}, \citenamefont
  {Psaila}, \citenamefont {Jose}, \citenamefont {Jha}, \citenamefont {Labadie},
  \citenamefont {Kern}, \citenamefont {Kar},\ and\ \citenamefont
  {Thomson}}]{Rodenas2012}%
  \BibitemOpen
  \bibfield  {author} {\bibinfo {author} {\bibfnamefont {A.}~\bibnamefont
  {R\'{o}denas}}, \bibinfo {author} {\bibfnamefont {G.}~\bibnamefont {Martin}},
  \bibinfo {author} {\bibfnamefont {B.}~\bibnamefont {Arezki}}, \bibinfo
  {author} {\bibfnamefont {N.}~\bibnamefont {Psaila}}, \bibinfo {author}
  {\bibfnamefont {G.}~\bibnamefont {Jose}}, \bibinfo {author} {\bibfnamefont
  {A.}~\bibnamefont {Jha}}, \bibinfo {author} {\bibfnamefont {L.}~\bibnamefont
  {Labadie}}, \bibinfo {author} {\bibfnamefont {P.}~\bibnamefont {Kern}},
  \bibinfo {author} {\bibfnamefont {A.}~\bibnamefont {Kar}}, \ and\ \bibinfo
  {author} {\bibfnamefont {R.}~\bibnamefont {Thomson}},\ }\href@noop {}
  {\bibfield  {journal} {\bibinfo  {journal} {Optics Letters}\ }\textbf
  {\bibinfo {volume} {37}},\ \bibinfo {pages} {392} (\bibinfo {year}
  {2012})}\BibitemShut {NoStop}%
\bibitem [{\citenamefont {Nasu}\ \emph {et~al.}(2005)\citenamefont {Nasu},
  \citenamefont {Kohtoku},\ and\ \citenamefont {Hibino}}]{Nasu2005}%
  \BibitemOpen
  \bibfield  {author} {\bibinfo {author} {\bibfnamefont {Y.}~\bibnamefont
  {Nasu}}, \bibinfo {author} {\bibfnamefont {M.}~\bibnamefont {Kohtoku}}, \
  and\ \bibinfo {author} {\bibfnamefont {Y.}~\bibnamefont {Hibino}},\ }\href
  {http://www.ncbi.nlm.nih.gov/pubmed/15832918} {\bibfield  {journal} {\bibinfo
   {journal} {Optics Letters}\ }\textbf {\bibinfo {volume} {30}},\ \bibinfo
  {pages} {723} (\bibinfo {year} {2005})}\BibitemShut {NoStop}%
\bibitem [{\citenamefont {Said}\ \emph {et~al.}(2004)\citenamefont {Said},
  \citenamefont {Dugan}, \citenamefont {Bado}, \citenamefont {Bellouard},
  \citenamefont {Scott},\ and\ \citenamefont {Mabesa}}]{Said2004}%
  \BibitemOpen
  \bibfield  {author} {\bibinfo {author} {\bibfnamefont {A.~A.}\ \bibnamefont
  {Said}}, \bibinfo {author} {\bibfnamefont {M.}~\bibnamefont {Dugan}},
  \bibinfo {author} {\bibfnamefont {P.}~\bibnamefont {Bado}}, \bibinfo {author}
  {\bibfnamefont {Y.}~\bibnamefont {Bellouard}}, \bibinfo {author}
  {\bibfnamefont {A.}~\bibnamefont {Scott}}, \ and\ \bibinfo {author}
  {\bibfnamefont {J.~R.}\ \bibnamefont {Mabesa}},\ }in\ \href {\doibase
  doi:10.1117/12.533540} {\emph {\bibinfo {booktitle} {Proc. SPIE 5339, Photon
  Processing in Microelectronics and Photonics III, 194}}}\ (\bibinfo {year}
  {2004})\BibitemShut {NoStop}%
\bibitem [{\citenamefont {Thomson}\ \emph {et~al.}(2011)\citenamefont
  {Thomson}, \citenamefont {Birks}, \citenamefont {Leon-Saval}, \citenamefont
  {Kar},\ and\ \citenamefont {Bland-Hawthorn}}]{Thomson2011}%
  \BibitemOpen
  \bibfield  {author} {\bibinfo {author} {\bibfnamefont {R.~R.}\ \bibnamefont
  {Thomson}}, \bibinfo {author} {\bibfnamefont {T.~A.}\ \bibnamefont {Birks}},
  \bibinfo {author} {\bibfnamefont {S.~G.}\ \bibnamefont {Leon-Saval}},
  \bibinfo {author} {\bibfnamefont {A.~K.}\ \bibnamefont {Kar}}, \ and\
  \bibinfo {author} {\bibfnamefont {J.}~\bibnamefont {Bland-Hawthorn}},\ }\href
  {http://www.ncbi.nlm.nih.gov/pubmed/21445210} {\bibfield  {journal} {\bibinfo
   {journal} {Optics Express}\ }\textbf {\bibinfo {volume} {19}},\ \bibinfo
  {pages} {5698} (\bibinfo {year} {2011})}\BibitemShut {NoStop}%
\bibitem [{\citenamefont {Eaton}\ \emph {et~al.}(2006)\citenamefont {Eaton},
  \citenamefont {Chen}, \citenamefont {Zhang}, \citenamefont {Zhang},
  \citenamefont {Iyer}, \citenamefont {Aitchison},\ and\ \citenamefont
  {Herman}}]{Eaton2006}%
  \BibitemOpen
  \bibfield  {author} {\bibinfo {author} {\bibfnamefont {S.~M.}\ \bibnamefont
  {Eaton}}, \bibinfo {author} {\bibfnamefont {W.}~\bibnamefont {Chen}},
  \bibinfo {author} {\bibfnamefont {L.}~\bibnamefont {Zhang}}, \bibinfo
  {author} {\bibfnamefont {H.}~\bibnamefont {Zhang}}, \bibinfo {author}
  {\bibfnamefont {R.}~\bibnamefont {Iyer}}, \bibinfo {author} {\bibfnamefont
  {J.~S.}\ \bibnamefont {Aitchison}}, \ and\ \bibinfo {author} {\bibfnamefont
  {P.~R.}\ \bibnamefont {Herman}},\ }\href@noop {} {\bibfield  {journal}
  {\bibinfo  {journal} {IEEE Photonics Technology Letters}\ }\textbf {\bibinfo
  {volume} {18}},\ \bibinfo {pages} {2174} (\bibinfo {year}
  {2006})}\BibitemShut {NoStop}%
\end{thebibliography}
\end{document}